\documentclass[11pt]{article}
\usepackage{vietnam,epsfig}

\def\be{\begin{equation}}
\def\ee{\end{equation}}
\def\bea{\begin{eqnarray}}
\def\eea{\end{eqnarray}}

\begin{document}
\vspace*{4cm}
\title{INTRODUCTION TO DARK ENERGY AND DARK MATTER}

\author{ PAUL H. FRAMPTON }

\address{Department of Physics and Astronomy, University of North Carolina,\\
Chapel Hill, NC 27599-3255}

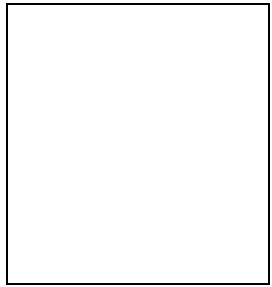
\maketitle\abstracts{
In an introductory manner, the nature of dark energy
is addressed, how it is observed and what further tests
are needed to reconstruct its properties. Several theoretical
approaches to dark energy will be discussed. Finally, the dark matter
especially WIMPs is introduced.
}

\section{Plan of Introduction to Dark Energy}
\begin{itemize}
\item What observations and theoretical assumptions underly dark energy (DE)?
\item If general relativity (GR) holds at all length scales, the most
conservative assumption, then DE follows from the supernovae Type 1A (SNe1A)
or, independently, from the Cosmic Microwave Background (CMB)
combined with Large Scale Structure (LSS).
\item Should we seriously query GR at large distance scales?

\end{itemize}

\section{Einstein-Friedmann Equation}

The Einstein equations relate geometry on the Left-Hand-Side (LHS)
to the distribution of mass-energy on the Right-Hand-Side (RHS)

\begin{equation}
G_{\mu\nu} = - 8 \pi G T_{\mu\nu}
\label{EinFried}
\end{equation}

We hesitate to change the LHS but it is really checked with precision
only at Solar System (SS) scales. At cosmological length scales, 
we may consider using a modification
such as higher-dimensional gravity.

On the RHS, if we include only luminous and dark matter it is
insufficient (keeping the LHS intact) and there is needed
a further term which could be a cosmological constant or, 
more generally, dark energy.

\section{Observational Issues}

How can we constrain DE?

\begin{itemize}

\item Measurement of the expansion history H(t)
\item The time-dependence of the equation of state w(t)
\item Looking for any clustering property of DE. No evidence for this presently.
\item How does DE couple to Dark Matter (DM)? This
is related to the question of clustering.
\item Local tests of GR and the equivalence principle, though the
extrapolation from the SS to the Universe is
some 13-15 orders of magnitude comparable to the
extrapolation from the
weak scale to the GUT scale in particle phenomenology.
The usual prior is a desert hypothesis.
\end{itemize}

\section{$\Lambda$ as DE: Why $10^{-122}$ (Planck Mass)$^4$?}

We know from the Lamb Shift and Casimir Effect in quantum
electrodynamics that vacuum fluctuations are 
real effects.

If we calculate the value of $\Lambda$, it will naively be
ultra-violet (UV) quartically divergent. The most natural UV cut-off in GR
is the Planck mass $\sim 10^{19} GeV$ whereupon
\begin{equation}
\Lambda \sim (10^{19}GeV)^4 = (10^{28} eV)^4 = 10^{112} (eV)^4
\end{equation}
If we use, instead, the weak scale $\sim 100 GeV$ as our UV cut-off, we arrive
at
\begin{equation}
\Lambda \sim (100 GeV)^4 = (10^{11} eV)^4 = 10^{44} (eV)^4
\end{equation}
The observed value for $\Lambda$, by contrast, is approximately
\begin{equation}
\Lambda \sim (3 \times 10^{-3} eV)^4 \sim  10^{-10} (eV)^4
\end{equation}

\section{Coincidence Problem}

As if the fine-tuning problem for $\Lambda$ were not enough, there
is a second problem with $\Lambda$, the coincidence problem.
Let us define $\Omega_{\Lambda} = \rho_{\Lambda}/\rho_{C}$
as the fraction of the critical density $\rho_{C}$.

The present value is $\Omega_{\Lambda} \sim 0.7$ but it scales,
since $\rho_{\Lambda}$ is constant and assuming $\Omega_{TOT} = 1$,
like $\rho_{C}^{-1} \sim (1 + Z)^{-3}$ so at a redshift $Z > 10$ it was
$\Omega_{\Lambda} < 0.001$ while for a future redshift $Z < -0.9$
one has $\Omega_{\Lambda} > 0.999$. 

If we plot $\Omega_{\Lambda}$ versus log R over cosmic history
from $ - 60 < log_{10} R < + 60$, it appears like a step function
changing from zero to one abruptly around the present era.
Even more dramatic is a plot of $ d\Omega_{\Lambda}/dR$
which approximates a Dirac delta function and the coincidence
problem is then why we live right in the middle of the spike of
the delta function.

If the dark energy had appeared earlier it would have interfered
with structure formation: if later, we would still be unaware
of it.

\section{The Quintessence Possibility}

One parametrization of the dark energy can be made using a 
dynamical scalar field, now generically called quintessence.

\subsection{Scaling potentials}

Examples are:

\begin{equation}
V \sim e^{-\lambda \Phi}
\end{equation}
as in \cite{W,FJ}
\begin{equation}
V \sim ((\Phi - A)^2 + C)e^{ - \lambda \Phi}
\end{equation}
as in \cite{AS}.

\subsection{Tracker Potentials}

Examples are 

\begin{equation}
V \sim \Phi^{- \alpha}
\end{equation}
as in \cite{RP},
\begin{equation}
V \sim exp \left( \frac{M}{Q} - 1 \right)
\end{equation}
as in \cite{SWZ}.

\subsection{Approaches to the Coincidence Problem}

We may assume that our universe sees periodic
epochs of acceleration\cite{DKS} with potential

\begin{equation}
V \sim M^4 e^{- \lambda \Phi} (1 + A \sin a \Phi)
\end{equation}

\bigskip

Another possibility is that it is important
that our epoch is close to the matter/radiation equality
time. This may be incorporated by having a non-minimal
coupling to matter\cite{BM}, to gravity\cite{PB}
or in a k-essence theory with a non-trivial
kinetic term in the lagrangian\cite{AP}.

\bigskip

\section{Dark Energy with Equation of State $w = p/\rho < -1$}

Present data on SNe1A, CMB and LSS are consistent with w=-1 as
for a cosmological constant.

Since the possibility that $w < -1$ is still allowed\cite{MMOT}, 
I shall spend a disproportionate amount of time on it because, if
it persisted, it could well signal new physics.

One interpretation of dark energy comes from string theory, closed strings
on a toroidal cosmology\cite{BFM}. This leads generically to
$w < -1$ \cite{F}.

In general, without dark energy (as in most cosmology texts pre-1998),
the destiny of the Universe was tied to geometry in
a simple manner: the Universe will expand forever
if it is open or flat; it will stop expanding and contract to a Big Crunch if it is closed.

With Dark Energy, this connection between geometry and destiny
is lost and the future fate depends entirely on how the
presently-dominant dark energy will evolve.

This question is studied in \cite{K,FT,RRC}. If $w < -1$ and
is time-independent, the scale
factor diverges at a finite future time
- the Big Rip.
Generally, this will be at least as far in the future as the
Big Bang was in the past.

Such a cosmology may have philosophical appeal? There is
more symmetry between past and future.

If one allows a time-dependent w(t), there are two other 
possible fates:

(i)An infinite-lifetime universe where dark
energy dominates at all future times.

(ii)A disappearing dark energy where the Universe
becomes (again) matter dominated.

The case $w < -1$ gives rise to some exceptionally interesting puzzles for
theoretical physics.

There is the question of violation of the weak energy condition
universally assumed in general relativity. This means
there are inertial frames where the energy density is negative
signaling vacuum instability\cite{F2004,CMT}.

Let us make three assumptions, any or all of which may be incorrect,
just so that we may say something more: that
(i) There is a stable vacuum with $\Lambda = 0$;
(ii) The dark energy decays to it by a 1st-order phase transition;
(iii) There is some, albeit feeble, interaction between dark energy
and the electromagnetic field.

Then one can use old arguments\cite{F1976} to investigate nucleation.
The result is that\cite{F2004} even with the tiniest coupling of dark energy to
the electromagnetic field the dark energy
would have spontaneously decayed long ago unless the appropriate
bubble radius is at least galactic in size. 

In this model, because the energy density of the DE is so small
compared to {\it e.g.} the energy
density in a common macroscopic magnetic field of, say, 10T
the 1st order phase transition can be adequately suppressed
only by decoupling the DE completely from
all but gravitational forces or by
arguing that a collision would need to be between
galaxies or larger objects to be effected.
Certainly, no terrestrial experiment can be influenced:
for one contrary suggestion of a Josephson junction
experiment which might well be justified for other reasons, 
see {\it e.g.} \cite{BeckMackey}.

Of course, this is only a toy model but the general
conclusion is probably correct - that there can
be no microscopic effect of the dark energy.

This makes the DE very difficult or impossible to
investigate except through
astronomical observations.

\section{Dark Energy and Neutrinos}

It has been pointed out by many theorists that
the density of the dark energy $\sim (10^{-3} eV)^4$
is suggestive of the neutrino mass.

Very interesting attempts to strengthen such a 
connection have been made
\cite{PQ,Nelson}. Such 
mass-varying neutrino models
seek to make a direct identification of the DE density
with neutrino mass\cite{F1,F2} itself.

\section{Precision Experiment}

We know well of the precision experiments to test Newton's
Law of Gravity down to a distance of 100 microns and below.

One originator of such ideas suggests\cite{DGZ} a different precision test,
of the Earth-Moon distance, to a similar accuracy of 100 microns, presumably
the distance between the centers of mass. A particular
modification of gravity\cite{DGP} might have a tiny
effect on our lunar system. Clearly if this experiment
can be achieved, the present accuracy being at the level of
centimeters, it would be an impressive achievement.

\section{Conclusions on Dark Energy}

\begin{itemize}

\item The theoretical community has yet to come up with a definitive proposal
to explain the dark energy.

\item The nature of the dark energy is so profound for cosmology
and particle physics that we desperately need more SNe1A observations from
important proposed experiments {\it e.g.} SNAP (for which NASA funding has sadly been
suspended for 5 years as a result of prioritizing sending humans to Mars!), as well
as complementary observational constraints on the CMB from {\it e.g.} the Planck mission.

\item The equation of state will be decisive. If w=-1, it's a cosmological
constant with its fine-tuning and coincidence problems. If $w > -1$
quintessence will receive a shot in the arm.

\item If the data would settle down to a value $w < -1$ we could be at the dawn
of a revolution in theory with general relativity at the largest distance scales
called into question.

\end{itemize}

\bigskip
\bigskip

\section{Introduction to Dark Matter}.

\bigskip

\noindent Existence of darl matter is supported by disparate cosmological measurements.

\bigskip

\noindent Values of energy and matter densities at the present time, determined
by: the temperature fluctuations in the CMB data; distance-luminosity for
supernovae type 1A; distribution of galaxies on large scales (LSS);
abundance of light elements (BBN).

\bigskip
\bigskip

\noindent In terms of the critical density
$\Omega$ for the various components is found to be as follows
(taking $h^2 = 0.5, h = 0.707$).

\bigskip

\begin{itemize}

\item Relativistic particles, radiation {\it e.g.} the
CMB photons. Only $\Omega_{\gamma} = 5.934 \pm 0.008 \times 10^{-5}$.

\bigskip

\item $\Omega_{\Lambda} = 0.72 \pm 0.08$ in a 
smoothly distributed dark energy.

\bigskip

\item $\Omega_{M} = 0.27 \pm 0.016$ in non-relativistic particles (Matter)
of which

\bigskip 

\noindent $\Omega_b = 0.0448 \pm 0.0018$ in baryons (protons and neutrons)

\bigskip

\noindent $\Omega_{HDM} < 0.0152 (95\% CL)$ in non-baryonic hot dark matter.

\bigskip

\noindent $\Omega_{CDM} = 0.223 \pm 0.016$ in non-baryonic cold dark matter.

\end{itemize}

\bigskip

\noindent The excess of total matter density ($0.27$) 
over baryonic mass density ($0.0224$) constitutes the evidence
for non-baryonic dark matter.

\noindent \underline{No known elementary particle can account for the non-baryonic dark matter}.

\bigskip

\noindent One obvious candidate, the neutrinos, are so light they constitute hot
dark matter and contribute to the $\Omega_{HDM} < 0.0152$.

\bigskip

\noindent Many hypothetical particles have been proposed for the CDM.
Some come from extensions of the standard model, most notably the
axion and the lightest supersymmetric particle.

\bigskip

\noindent Other possibilities include Wimpzillas, solitons, self-interacting dark matter,
Kaluza-Klein dark matter, etc.

\bigskip

\noindent The class of non-baryonic dark matter candidate
of greatest interest are the Weakly Interacting Massive Particles (WIMPs).
Therefore I shall focus on their detection and the claims to have discovered them.

\bigskip
\bigskip

\section{WIMPs and their detection}

\bigskip

\noindent WIMPs are appealing because of the simple mechanism by which they
can achieve the appropriate present cosmic density. In the early
universe they were in thermal and chemical equilibrium with the rest
of matter and radiation. With the expansion of the Universe, their
reactions (including annihilation) slowed down and decoupled from
the rest of the world leaving a constant number of WIMPs expanding
with the Universe.

\bigskip

\noindent The correct present density is obtained for WIMPs with couplings
of order the weak interactions and masses in the 1 GeV - 1 TeV range.
The neutralino\cite{Goldberg} is the most popular example although
in any extension of the standard model one typical seeks a WIMP
candidate, {\it e.g.} the nark is a WIMP candidate\cite{AFNNY} in the sark model\cite{FN}.

\noindent Detection can be direct or indirect.

\bigskip

\noindent \underline{Direct} signals are from collisions with
nuclei in a detector. A very sensitive low-background detector
(bolometer) records the amount of energy deposited by WIMPs
and (in the future) the direction of motion of the struck nucleus.

\bigskip

\noindent \underline{Indirect} signals come from WIMP reactions in 
planets, stars or galaxies. The most common reaction is
WIMP annihilation with anti-WIMP. Out of this annihilation come
$\nu$, $e^+$, $\bar{p}$ and high-energy $\gamma$.
Such annihilations occur at a detectable rate where the anti-WIMPs
are concentrated {\it e.g.} in the center of
the Sun, the center of the Earth and in galactic centers 
including the Milky Way. 
Neutrino telescopes, gamma-ray telescopes and cosmic-ray detectors
can be used in these indirect searches.

\bigskip

Now we look at three claims for seeing WIMPs.

\bigskip
\bigskip

\section{HEAT positron detection}

\bigskip

\noindent Two separate balloon flights with different detectors
have seen\cite{HEAT} more cosmic ray positrons above $\sim 7$ GeV
than predicted in models for
cosmic ray propagation in the galaxy.
Wimp annihilation can be invoked to explain this excess.

\bigskip

The extra positrons
can be fitted by a assuming neutralino annihilation. The best fit
requires
WIMP mass of 238 GeV.

\bigskip

The positron spectrum lacks any
discriminating feature which clearly
singles out WIMP annihilation.

\bigskip
\bigskip

\section{$\gamma$-Rays from Galactic Bulge}

\bigskip

Gamma rays from WIMP annihilation offer a characteristic
signature in the spectrum: a gamma-ray line.
Each photon will carry an energy equal to the WIMP mass,
10GeV to 100 TeV. No competing process is know that could produce
such a line.

\bigskip

No line has been detected yet. The estimates for GLAST (launch scheduled 2006)
are encouraging.

\bigskip

Another suggestion\cite{silk} has been that the 511 keV gamma excess from the
galactic bulge arises from positrons associated with unexpectedly
light WIMP annihilation. The necessary WIMP mass is in the region
between 1 MeV and 100 MeV. There are other explanations for the 511 KeV line
including primordial black holes as dark matter\cite{FKPBH}.

\bigskip
\bigskip

\section{DAMA Annular Modulation}

\bigskip

Because the Earth's motion changes the relative speed of the Earth
and WIMPs the WIMP detection rate varies\cite{drukier} and repeats itself
every year. The maximum occurs in June for the canonical halo
model with Maxwellian velocity distribution.

\bigskip

The DAMA group has claimed\cite{DAMA} to have detected such annual modulation in
their NaI data. No alternative explanation of the DAMA data has
been forthcoming.

\bigskip

\noindent No other direct detection of such a WIMP signal
has been made by any other group
but there are differences between the targets used
as well as the nuclear spin thereof. So comparison between experiments
requires some theoretical assumptions.

\bigskip

\noindent Nevertheless, it does appear that CDMS data\cite{CDMS} {\it completely} 
or {\it almost (?)} excludes the DAMA claim.

\bigskip

Future detectors will measure the direction of motion of the
recoil nucleus and enable a more clearcut WIMP signature.

\bigskip
\bigskip

\section{Conclusions on  DARK MATTER}

\bigskip

\noindent How to be Sure of WIMP Detection?

\bigskip

\noindent We require features that can be due to WIMPs and nothing else.

\bigskip

\begin{itemize}

\bigskip

\item (i) Gamma-ray annihilation from WIMP annihilation
should show a gamma-line in correspondence with the WIMP mass.

\bigskip

\item (ii)Annual modulation should show the correct periodicity both in rate and, in future,
directional dependence.

\end{itemize}

\bigskip 

\noindent Compatible indirect (i) and direct (ii) detection 
could provide compelling evidence for WIMPs.

\bigskip

\noindent Better would be production in a collider consistent with cosmological detection!

\bigskip
\bigskip

\section*{Acknowledgments}
We thank Tran Thanh Van for the invitation to La Thuile.
This work was supported in part by the US Department of Energy
under Grant No. DE-FG02-97ER-41036.

\end{document}